\documentclass[reprint,aps,superscriptaddress,amsmath,amssymb]{revtex4-2}
\usepackage{graphicx}
\usepackage[linkcolor=blue,urlcolor=blue,citecolor=blue,colorlinks=true]{hyperref}

\def\DTR{\Delta\tau_\textsc{r}}
\def\DTW{\Delta\tau_\textsc{ews}}
\def\TTT{\tau_\textsc{t}}
\def\TTR{\tau_\textsc{r}}

\def\VI{V_{\rm i}}
\def\BR{a_0}

\def\ETAT{\eta_{\textsc t}}
\def\ETAR{\eta_{\textsc r}}

\begin{document}
\title{Attosecond dynamics of electron scattering by an absorbing layer}
\author{R.O. Kuzian}
\affiliation{Donostia International Physics Center (DIPC), 20018 Donostia/San Sebasti\'{a}n, Basque Country, Spain}
\affiliation{Institute for Problems of Materials Science NASU, Krzhizhanovskogo 3, 03180 Kiev, Ukraine}
\author{E. E. Krasovskii}
\affiliation{Donostia International Physics Center (DIPC), 20018 Donostia/San Sebasti\'{a}n, Basque Country, Spain}
\affiliation{Universidad del Pais Vasco/Euskal Herriko Unibertsitatea, 20080 Donostia/San Sebasti\'{a}n, Basque Country, Spain}
\affiliation{IKERBASQUE, Basque Foundation for Science, 48013 Bilbao, Basque Country, Spain}

\begin{abstract}
Attosecond dynamics of electron reflection from a thin film is studied 
based on a
one-dimensional jellium model. Following the Eisenbud-Wigner-Smith concept, 
the reflection time
delay $\DTR$ is calculated as the energy derivative of the phase of the 
complex reflection
amplitude $r$. For a purely elastic scattering by a jellium slab of a finite thickness $d$
the transmission probability $T$ oscillates with the momentum $K$ in the solid with a period
$\pi/d$, and $\DTR$ closely follows these oscillations. The reflection delay averaged over an
energy interval grows with $d$, but in the limit of $d\to\infty$ the
amplitude $r$ becomes real, so $\DTR$ vanishes. This picture changes substantially with the
inclusion of an absorbing potential $-i\VI$: As expected, for a sufficiently thick slab the
reflection amplitude now tends to its asymptotic value for a semi-infinite crystal.
Interestingly, for $\VI\ne 0$, around the $T(E)$ maxima, the $\DTR(E)$ curve strongly deviates
from $T(E)$, showing a narrow dip just at the $\DTR(E)$ maximum for $\VI=0$. An analytical
theory of this counterintuitive behavior is developed. 
\end{abstract}

\maketitle

\section{Introduction}

Rapid progress in time-resolved electron spectroscopy, in particular in the experiments on
attosecond streaking~\cite{Cavalieri2007,Schultze10,Neppl2012,Neppl15,Okell15,Siek2017,
Ossiander2018,Riemensberger2019} and RABBITT interferometry~\cite{Locher:15,Tao2016},
has drawn the attention to microscopic theoretical approaches capable of underlying
{\it ab initio} methods applicable to realistic solids. In contrast to 
stationary photoemission,
where a steady flux of light causes steady photocurrent, the photoelectron excited by a short
light pulse is naturally viewed as a wave packet that propagates through the medium toward
the detector. One fruitful approach to the propagation timing stems from the concept of
Eisenbud-Wigner-Smith time delay~\cite{Bohm1951,Wigner55,Smith60}, which allows to express the
asymptotic equation of motion in terms of stationary states, namely as the energy derivative
of the scattering phase.

Consider a wave packet composed of plane waves with the wave vectors along $\mathbf z$: For a
finite scatterer, the maximum of the incident wave packet asymptotically far from the scatterer
has the equation of motion $z_{\textsc i}(t)=vt$, where $v=\hbar k/m$ is the group velocity of
the free motion. Behind the scatterer, the transmitted wave packet obeys the asymptotic
equation of motion
\begin{equation}
z_{\textsc t}(t)=v(t-\DTW), \label{eq:eomscat}
\end{equation}
where $\DTW$ is known as the phase time or Eisenbud-Wigner-Smith (EWS) time delay, and
for a spectrally narrow wave packet it equals the energy derivative of the scattering
phase $\ETAT$~\cite{Bohm1951,Wigner55,Smith60}:
\begin{equation}
\DTW = \hbar d\ETAT/dE\equiv\dot{\eta}. \label{eq:EWS}
\end{equation}

If the starting and the ending points are sufficiently far away from the scattering region
then $\DTW$ can be interpreted as the difference between the time needed by a free electron
to transit the distance between the two points and the actual time needed by the scattered
electron. The problem of transit time has been addressed in numerous studies, with main
attention paid to paradoxes related to elastic scattering, see review
articles~\cite{Hauge1989,Landauer1994,Winful2006,Field2022}.

The implications of the inelastic scattering remain less studied despite its importance
in electron spectroscopies: Electron damping leads to a finite electron escape depth in
photoemission~\cite{FeibelmanEastman_1974} and decreases reflectivity in the low energy
electron diffraction (LEED) experiment~\cite{Slater37}. The standard method to take into
account electron damping used since it was suggested by Slater in 1937~\cite{Slater37}
consists in adding am imaginary potential $-i\VI$ to the  crystal potential but keeping
the energy $E$ real in the Schr\"odinger equation $\hat H\Psi=E\Psi$. (Alternatively,
non-hermiticity can be introduced by nonreciprocal hopping in the tight-binding
picture~\cite{Longhi2022}.) The optical potential $-i\VI$ is associated with the imaginary
part of the electron self-energy~\cite{Jepsen1982,Krasovskii2016} due to the interaction
with the electron environment. This phenomenological description is commonly used in the
one-step theory of the stationary photoemission~\cite{Pendry76,Braun96} to simulate the
surface sensitivity of the measurement and in the theory of LEED to reproduce the shape
of the experimental reflection and transmission spectra~\cite{Krasovskii_2002}. Furthermore,
the $\VI$ formalism was applied to electron propagation~\cite{Delgado2004} and time-resolved
photoemission~\cite{Krasovskii2016}. In Ref.~\cite{Krasovskii2016} the results by a
time-dependent Schr\"odinger equation with absorbing potential were compared to the inelastic
scattering treated fully microscopically by means of random collisions. 
In the former approach the dephasing of the wave packet is neglected, while in the 
latter the observables are obtained from a statistical averaging over 
random perturbations. The two schemes were found to agree well regarding 
the delay time in a numerical streaking experiment.

This makes it quite reasonable to try to take advantage of the artificial coherence of the
scattering state and apply the EWS formalism also in the presence of the 
inelastic scattering.
Here, we consider transmission and reflection of a wave packet incident on an absorptive
potential well within a one-dimensional (1D) jellium model. This problem 
has an exact analytical
solution, which allows to unambiguously establish a rather nontrivial effect of the optical
potential on the energy dependence of reflection time delay: at the points of complete
transparency on switching on the absorption the $\DTW(E)$ maxima turn into sharp 
local minima,
which become broader and more shallow with increase of $\VI$. 

The paper is organized as follows: The model is introduced in Sec.~\ref{sec:Model}. 
Section~\ref{sec:Real-potential} analyses time delays in the absence of the inelastic
scattering. The influence of the absorbing potential is discussed in
Sec.~\ref{sec:Absorbing-pot} followed by conclusions in Sec.~\ref{sec:Conclusion}.
The details of formulae derivation and some explicit expressions are presented
in~\ref{sec:Appendix}.

\section{Model}\label{sec:Model}
For a normal incidence on a jellium slab the problem reduces to a textbook 1D problem
of a particle scattered by a rectangular potential well. Let the well extend from $z=0$
to $z=d$, so the Hamiltonian in atomic units $\hbar=e=m=1$ is
\begin{flalign}
\hat{H} & =-\frac{1}{2}\frac{\partial^{2}}{\partial z^{2}}+V(z),\label{eq:H}
\end{flalign}
with the potential
\begin{flalign}
V(z) & =\begin{cases}
-U, & 0\leq z\leq d,\\
\hfill 0, & \mathrm{otherwise}.
\end{cases}\label{eq:Vz}
\end{flalign}

The space is divided into tree regions: the half-space to the left of the slab, a finite
scattering region, and the half-space to the right of the slab. The solution of
the stationary Schr\"odinger equation $\hat{H}\psi=E\psi$ for a wave incident from the left at
a positive energy $E=k^{2}/2$ is
\begin{equation}
\psi(z)=\begin{cases}
\hfill e^{ikz}+re^{-ikz}, & z<0,\\
Ae^{iKz}+Be^{-iKz}, & 0\leq z\leq d,\\
\hfill te^{ikz}, & z>d.
\end{cases}\label{eq:psi}
\end{equation}
In the left half-space the wave function consists of the incident wave of unit amplitude
and a reflected wave moving in the opposite direction with the amplitude
$r=i\left|r\right|\exp(i\ETAR)$. In the scattering region, it is a sum of two waves
propagating in opposite directions. For a potential well ($U>0$), $K$ is larger than the
free-space wave vector $k$. (For a scatterer that is a fragment of a periodic potential
these are two Bloch waves with the Bloch vectors $K$ and $-K$.) In the right half-space,
there is only the transmitted wave moving to the right with the amplitude
$t=\left|t\right|\exp(i\ETAT)$.

The amplitudes are found from the condition of the continuity of the function $\psi(z)$
and its derivative $\partial\psi/\partial z$ at the boundaries $z=0$ and $z=d$,
see~\ref{sec:Appendix}:
\begin{flalign}
& t=\frac{e^{-ikd}}{\cos\left(Kd\right)-i\dfrac{K^{2}+k^{2}}{2kK}\sin\left(Kd\right)},
\label{eq:t}\\
 & r=i\frac{K^{2}-k^{2}}{2kK}\sin\left(Kd\right)te^{ikd},\label{eq:r}\\
 & A=\frac{K+k}{2K}te^{i\left(k-K\right)d}=\frac{K+k+r\left(K-k\right)}{2K},\label{eq:A}\\
 & B=\frac{K-k}{2K}te^{i\left(k+K\right)d}=\frac{K-k+r\left(K+k\right)}{2K}.\label{eq:B}
\end{flalign}

From the stationary wave function of Eq.~(\ref{eq:psi}) we now obtain the EWS
time delay of the transmitted and reflected wave packets as the energy derivatives
of the transmission and reflection phases $\ETAT$ and $\ETAR$ derived from
Eqs.~(\ref{eq:t}) and~(\ref{eq:r}), respectively.

\section{Purely elastic scattering\label{sec:Real-potential}}

First, let us consider a real potential $U=V_{0}$. Then the wave vectors in the
scattering region are also real, $K=\sqrt{2(E+V_{0})}$, and Eq.~(\ref{eq:r})
yields the simple relation between the phases of transmission and reflection
amplitudes~\cite{Falck1988}
\begin{equation}
\ETAR=\frac{\pi}{2}+\ETAT+kd. \label{eq:etareal}
\end{equation}
Similar to transmission, Eq.~(\ref{eq:EWS}), the time delay on reflection $\Delta\TTR$ equals
the energy derivative of the phase shift $\ETAR$ of the reflected wave,
$\Delta\TTR =\dot{\ETAR}$.

Following Hartman~\cite{Hartman1962} we also introduce the so-called transit
time $\TTT=\DTW+d/k$, which is the time a classical particle would spend in
the scattering region if it obeyed the equations of motion (\ref{eq:eomscat}).
Owing to the relation~(\ref{eq:etareal}), the reflection time delay coincides
with the transit time (recall that $v=k$ in the atomic units):
\begin{equation}
\DTR=\DTW+d/k.\label{eq:tauReal}
\end{equation}
The transmission amplitude $t$ in Eq.~(\ref{eq:t}) has the form 
\begin{equation}
te^{ikd}=\frac{1}{a-ib},\label{eq:teikd}
\end{equation}
where real and imaginary parts of the denominator are 
\begin{equation}
a=\cos\left(Kd\right),\quad b=\frac{K^{2}+k^{2}}{2kK}\sin\left(Kd\right).
\label{eq:abreal}
\end{equation}
Then the transmission probability $T\equiv|t|^{2}$ is 
\begin{align}
T =\frac{1}{a^{2}+b^{2}} 
 =\frac{1}{1+\left[\dfrac{V_{0}}{kK}\sin(Kd)\right]^{2}}.\label{eq:modt2real}
\end{align}
Equation~(\ref{eq:modt2real}) shows that the transmission probability $T$
oscillates as a function of $K$ with the period $\pi/d$: There is a finite
probability to be reflected from the potential well at all energies except
for the transparency resonances. The $T=1$ resonances occur at
$K_{n}^{\mathrm{res}}=\pi n/d$, where the integer $n$ is greater than
$d\sqrt{2V_{0}}/\pi$, i.e., at the energies $E_{n}^{\mathrm{res}}=(\pi n/d)^{2}/2-V_{0}$.
The phase of the transmission amplitude is then 
\begin{flalign}
\ETAT +kd=\arctan\frac{d}{a}=\arctan\left[\frac{k^{2}+V_{0}}{kK}\tan(Kd)\right],
\nonumber 
\end{flalign}
and the transit time is
\begin{align}
\TTT & =\dot{\ETAT}+d/k=|t|^{2}(a\dot{b}-b\dot{a})\nonumber \\
& =\frac{T}{kK^{2}}\left[d\left(k^{2}+V_{0}\right)-V_{0}^{2}\frac{\sin(2Kd)}{k^{2}K}\right].
\label{eq:tauTreal}
\end{align}
\begin{figure}  
\centering 
\includegraphics[width=\columnwidth]{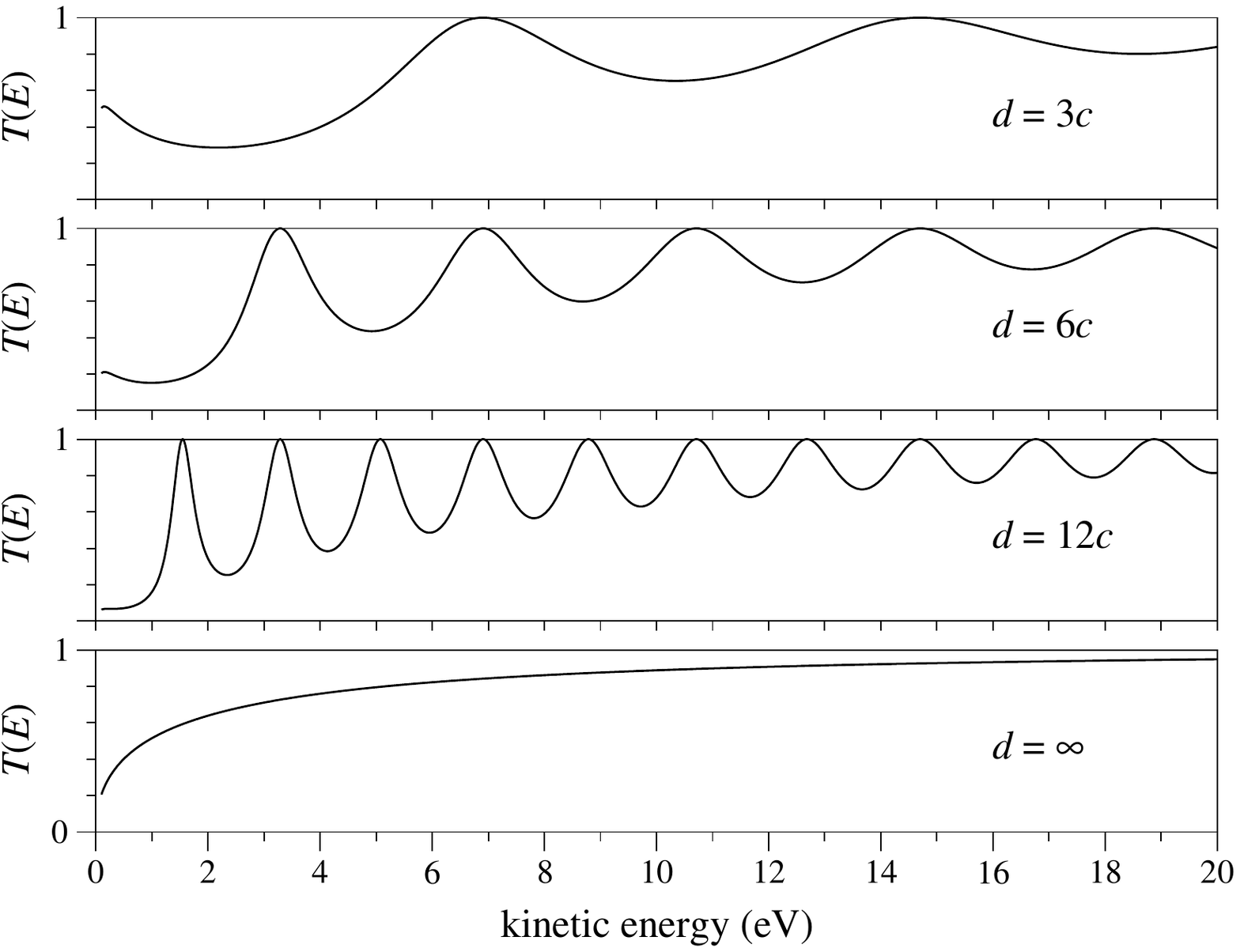}
\caption{Energy dependence of the transmission probability $T=|t|^{2}$ for
  potential wells of different widths, $d=nc$ with $c=6.34\:\BR$. The potential
  inside the well is $-U=V_{0}=-30.21$~eV. The period of oscillations is $\pi/d$.}
\label{fig1} 
\end{figure} 

\begin{figure} 
\centering 
\includegraphics[width=\columnwidth]{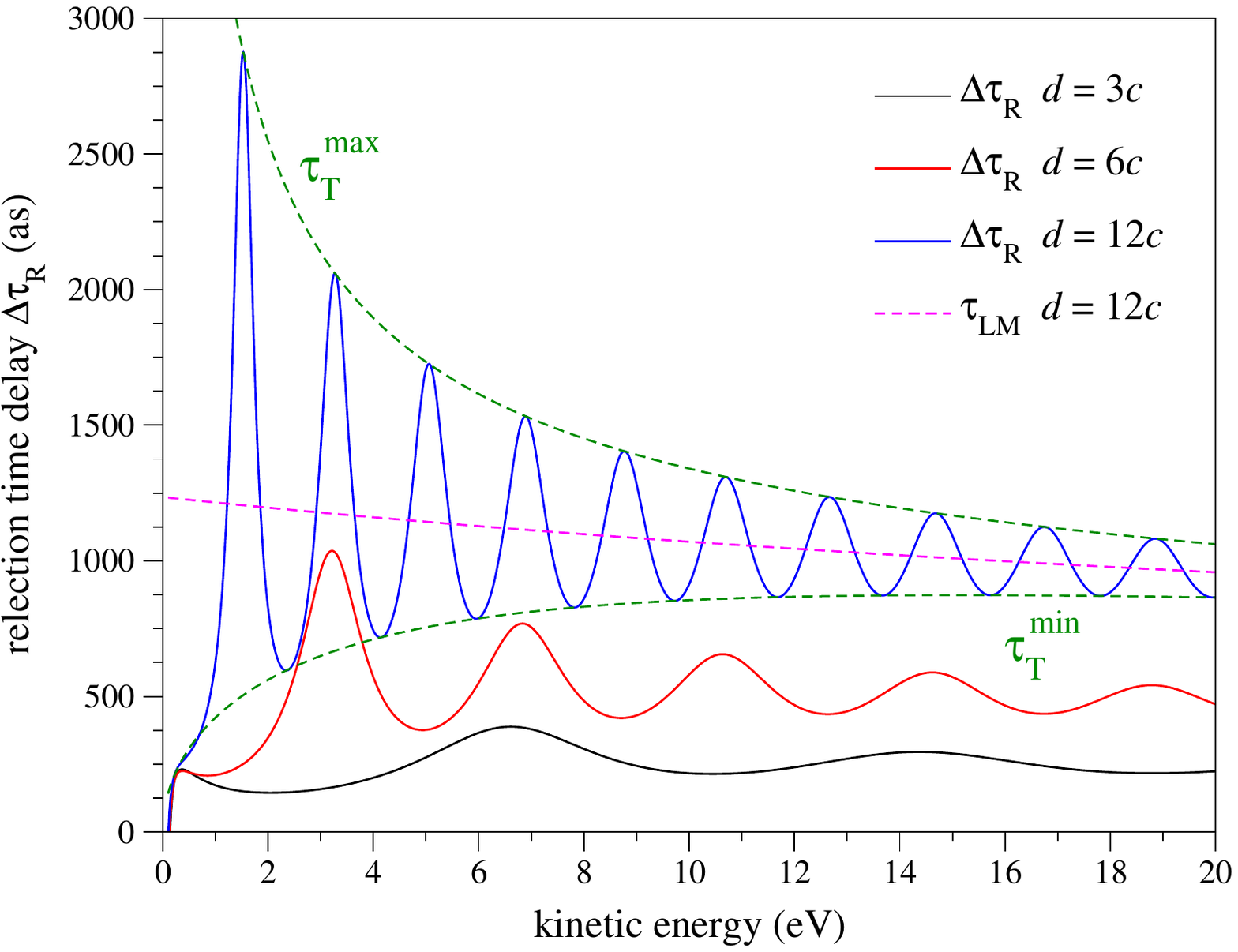}
\caption{Reflection time delay $\Delta\TTR(E)=\TTT(E)$ for three widths of the
potential well, $d=3c$, 6$c$, and 12$c$ (solid lines). The energy locations of
the maxima are close to the those of the $T(E)$ maxima, $E_{n}^{\mathrm{res}}$.
Dashed lines show the envelope functions $\TTT^{\mathrm{min}}$ and $\TTT^{\mathrm{max}}$
and the Landauer-Martin formula $\tau_{\textsc{lm}}=d/v_{g}$, see Eq.~(6.2) in
Ref.~\cite{Landauer1994}.
}
\label{fig2} 
\end{figure}

Figure~\ref{fig1} shows the energy dependence of the transmission probability
$T(E)$ for the scattering regions of different widths $d$. We have chosen the
parameters that roughly model the graphene multilayers studied in
Refs.~\cite{Jobst16,Krasovskii21,Krasovskii22,Krasovskii2024}, namely
$c=6.34~\BR$ and $V_{0}=30.21$~eV. According to Eq.~(\ref{eq:tauTreal}),
the transit time oscillates between the two envelope functions 
\begin{align}
\TTT^{\mathrm{max}} & =d\frac{(k^{2}+V_{0})}{kK^{2}}\;\text{ and }\;\label{eq:dtmax}\\
\TTT^{\mathrm{min}} & =d\frac{k(k^{2}+V_{0})}{(kK)^{2}+V_{0}^{2}}\label{eq:dtmin}
\end{align}
as a result of the multiple reflections from both boundaries of the well~\cite{Bohm1951}.
The same behavior for the reflection time delay follows from Eq.~(\ref{eq:tauReal}), see
Fig.~\ref{fig2}. 

Landauer and Martin~\cite{Landauer1994} proposed an approximate formula for the
transit time $\tau_{\textsc{lm}}=d/v_{g}$, where $v_{g}$ is the group velocity in
the crystal. In our model, $v_{g}=K$, and $\tau_{\textsc{lm}}$ lies between the
envelope functions $\TTT^{\mathrm{min}}$ and $\TTT^{\mathrm{max}}$. Let us recall that
Eq.~(\ref{eq:tauTreal}) gives the transit time for the wave packets having the
spectral width much smaller than the distance between the maxima of $\TTT$, i.e., the
packets composed of the waves with the energies
$|E-E_{n}^{\mathrm{res}}|\ll(\pi/d)^{2}(n+1/2)$. One may suppose
that the Landauer-Martin formula gives an average time delay for a packet, whose
spectral width is larger than the energy distance between the neighboring maxima
$E_{n}^{\mathrm{res}}$. Such a packet may be represented as a sum of spectrally narrow
packets, which are transmitted (and reflected) with substantially different delays.
Thus, the sum of the transmitted narrow packets have the shape substantially
different from the incident packet, and it may even split into several packets. The
reflected packet is similarly distorted. Thus, the Landauer-Martin formula cannot be
unambiguously interpreted as an average transit time for a spectrally wide packet.

We see from expressions (\ref{eq:dtmax}) and (\ref{eq:dtmin}) that the upper and
lower limits of $\TTT$ are proportional to the slab width $d$, and according to
Eq.~(\ref{eq:tauTreal}) $\TTT$ increases with the width. Now let us consider a slab
of infinite width, $d\to\infty$, i.e., the reflection from a semi-infinite crystal,
which in our model means the reflection from a step function. As shown in
\ref{sec:Appendix}, the reflection amplitude 
\begin{equation}
r_{\infty}=\frac{k-K}{k+K} \label{eq:rinfreal}
\end{equation}
in this case is \emph{real} and, consequently, the reflection time delay
\emph{vanishes} instead of the intuitively expected divergence with $d\to\infty$.
So, the reflected wave packet conserves shape and is a reduced copy of the incident
packet, with the equation of motion for its maximum being the same as for the
ideally reflected packet, $z_{\textsc r}(t)=-vt$. This is a consequence of the
absence of the second boundary: there is no interference with the internally reflected
waves.

\section{Inclusion of the absorbing potential\label{sec:Absorbing-pot}}
For a realistic description of scattering, the inelastic processes must be
included, to which end we add an imaginary part to the potential in the
scatterer: $U=V_{0}+i\VI$, see Eqs.~(\ref{eq:H}) and (\ref{eq:Vz}). Equations
(\ref{eq:t})--(\ref{eq:B}) for the coefficients of the partial waves remain
the same, but the wave vector in the slab becomes complex $K=K_{1}+iK_{2}$.
It satisfies the equation 
\begin{equation}
K^{2}=2(E+V_{0}+i\VI), \label{eq:K2}
\end{equation}
which gives 
\begin{align}
K_{1} & =\sqrt{\sqrt{\left(E+V_{0}\right)^{2}+\VI^{2}}+E+V_{0}}, \label{eq:K_1}\\
K_{2} & =\frac{\VI}{K_{1}}.\label{eq:K_2}
\end{align}
Substituting the complex vector $K$ into Eq.~(\ref{eq:t}) we find the
transmission amplitude in the form similar to Eq.~(\ref{eq:teikd}):
\begin{equation}
te^{ikd}=\frac{1}{p-iq}. \label{eq:teikdVi}
\end{equation}
The expressions for $p$ and $q$ are more involved than
Eq.~(\ref{eq:abreal}), and they are given in \ref{sec:Appendix},
Eqs.~(\ref{eq:at}) and (\ref{eq:bt}).

Similar to Eq.~(\ref{eq:tauTreal}) the transit time is 
\begin{equation}
\TTT=|t|^{2} (\dot{q}p-q\dot{p} ).
\label{eq:etatdot}
\end{equation}
Explicit expressions for $\dot{p}$ and $\dot{q}$ are given in
Eqs.~(\ref{eq:atdot}) and~(\ref{eq:btdot}) in \ref{sec:Appendix}.

\begin{figure} 
\centering 
\includegraphics[width=\columnwidth]{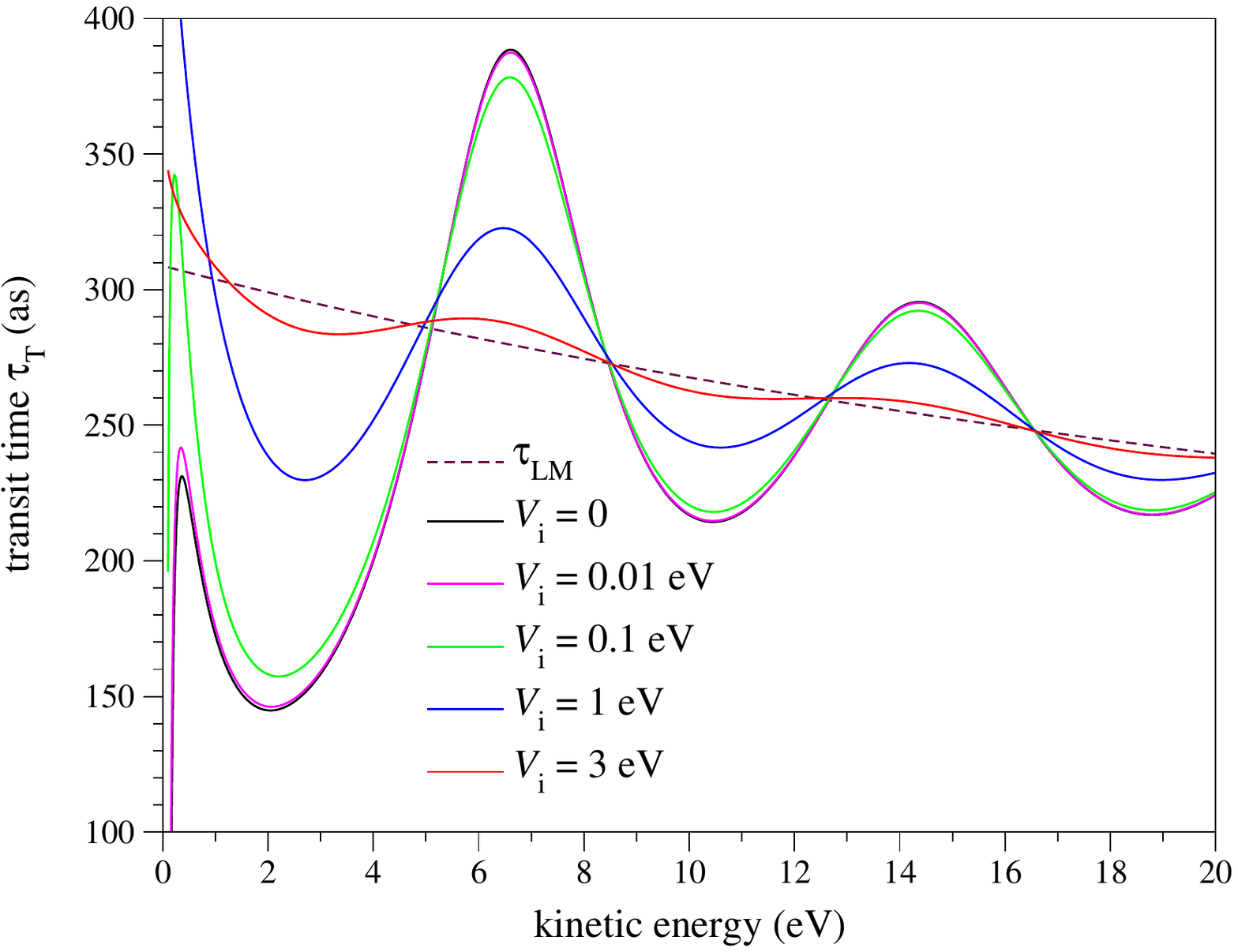}
\caption{Transit time $\TTT$ for a slab width $d=3c$ for five values of the
absorbing potential $\VI=0$, 0.01, 0.1, 1, and 3~eV (solid lines). The real
part of the potential is $V_{0}=30.21$~eV, as in Figs.~\ref{fig1} and~\ref{fig2}.
Dashed curve is the Landauer-Martin time $\tau_{\textsc{lm}}=d/v_{g}$.
}
\label{fig3} 
\end{figure}   

Figure~\ref{fig3} shows the transit time for different values of $\VI$
for the scatterer width $d=3c$. The absorbing potential is seen to damp
the oscillations of $\TTT$, which is an expected result of the inclusion
of absorbing potential: The $-i\VI$ term leads to a finite imaginary part
of the wave vector $K$ inside the slab, whereby the wave function decays
into the solid, and the wave incident from the left becomes less sensitive
to the presence of the right boundary of the slab.  Note a qualitative
difference of the effect of the wave function decay due to absorption from
the decay due to tunneling: In the latter case the transit time becomes
independent on the width $d$ leading to nominally unlimited propagation
velocities---the famous Hartman effect~\cite{Hartman1962}. 

\begin{figure}[t] 
\centering 
\includegraphics[width=\columnwidth]{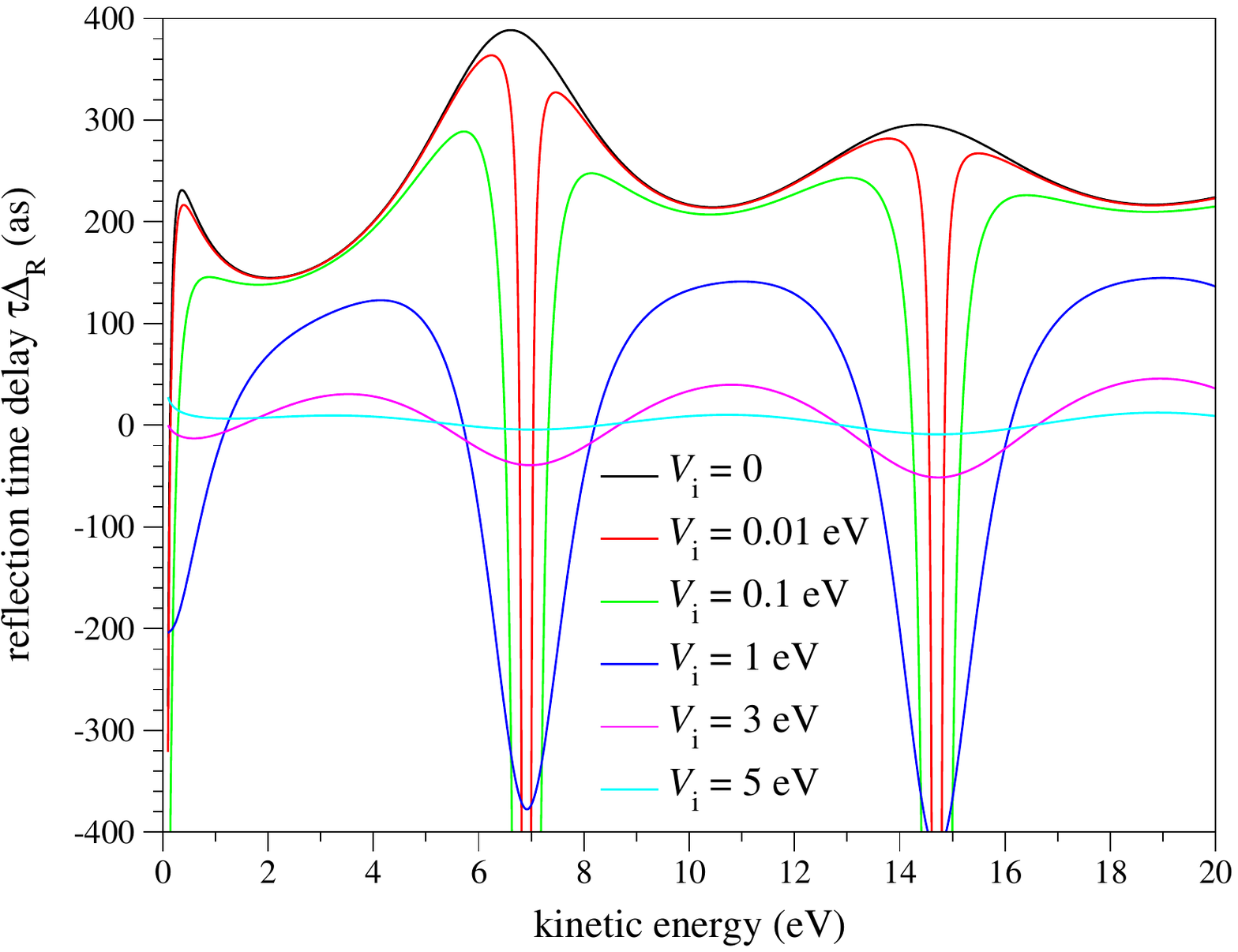}
\caption{Reflection time delay $\TTR$ for $d=3c$ and six values of
absorbing potential $\VI=0,\:0.01,\:0.1,\:1,\:3$, and 5~eV.
}
\label{fig4} 
\end{figure}

Let us now consider reflection. For a perfectly reflecting impenetrable plane
at $z=0$ the equation of motion for the maximum of the reflected wave packet is
$z_{\textsc r}(t)=-vt$, and the scatterer is characterized by the reflection time delay
$\Delta\TTR$ that modifies the equation of motion: $z_{\textsc r}(t)=-v(t-\Delta\TTR)$.
The effect of $\VI$ on the timing of the reflected wave packet is rather
unexpected. In contrast to the elastic case, for a finite $\VI$ the reflection
time delay $\DTR$ does not equal the transit time. Equation~(\ref{eq:tauReal}) does
not hold because the relation~(\ref{eq:etareal}) between the reflection and
transmission phases is valid only for real $K$. The reason is the energy
dependence of the imaginary part of $K$, see Eq.~(\ref{eq:K_2}), which results in
the energy dependence of the phases of all the terms in Eq.~(\ref{eq:r}). Thus,
for a finite $\VI$, instead of Eq.~(\ref{eq:etareal}) we have
\begin{equation}
\ETAR=\frac{\pi}{2}+\eta_{1}+\eta_{2}+\ETAT+kd,\label{eq:etar}
\end{equation}
where 
\begin{align}
\eta_{1} & =\mathrm{arg}\frac{K^{2}-k^{2}}{2kK}=
\arctan\frac{b_{1}}{a_{1}},\label{eq:eta1}\\
a_{1} & =K_{1}(|K|^{2}-k^{2}),\nonumber \\
b_{1} & =K_{2}(|K|^{2}+k^{2})\nonumber \\
\eta_{2} & =\mathrm{arg}\left[\sin\left(Kd\right)\right]=
\arctan\frac{b_{2}}{a_{2}}\label{eq:eta2}\\
a_{2} & =\tan(K_{1}d)\nonumber \\
b_{2} & =\tanh(K_{2}d).\nonumber 
\end{align}
Then, the reflection delay $\Delta\TTR$ is the sum of
the transit time $\TTT$ and two additional terms
\begin{equation}
\Delta\TTR=\dot{\eta}_{1}+\dot{\eta}_{2}+\TTT, \label{eq:tauWr}
\end{equation}
where 
\begin{align*}
\dot{\eta}_{i} & =\frac{a_{i}\dot{b}_{i}-b_{i}\dot{a}_{i}}{a_{i}^{2}+b_{i}^{2}},\\
\dot{a}_{1} & =(3K_{1}^{2}+K_{2}^{2}-k^{2})\dot{K}_{1}+2K_{1}(K_{2}\dot{K}_{2}-1),\\
\dot{b}_{1} & =(3K_{2}^{2}+K_{1}^{2}+k^{2})\dot{K}_{2}+2K_{2}(K_{1}\dot{K}_{1}+1),\\
\dot{a}_{2} & =\frac{d\dot{K}_{1}}{\cos^{2}\left(K_{1}d\right)},\\
\dot{b}_{2} & =\frac{d\dot{K}_{2}}{\cosh^{2}\left(K_{2}d\right)}.
\end{align*}
The delay $\dot{\eta}_{1}$ does not depend on the width $d$ and becomes small
compared to the transit time with the increase of the width. The delay
$\dot{\eta}_{2}$ is comparable to the transit time because it is proportional
to $d$. It is instructive to expand the delay near a point of complete
transparency $K_{n}^{\mathrm{res}}=\pi n/d$ for $\VI=0$. We set
$K_{1}=K_{n}^{\mathrm{res}}+\Delta K$ with $\Delta Kd\ll1$ and assume, quite
realistically, that $\VI/V_{0}\ll1$. Then 
\begin{align}
\dot{\eta}_{2} & =\frac{d}{\tan^{2}(K_{1}d)+\tanh^{2}(K_{2}d)}
\left[\frac{\tan(K_{1}d)\dot{K}_{2}}{\cosh^{2}\left(K_{2}d\right)}\right. \nonumber \\
 & \left. -\frac{\tanh(K_{2}d)\dot{K}_{1}}{\cos^{2}\left(K_{1}d\right)}\right]
 \approx-\frac{V_{i}}{(K_{n}^{\mathrm{res}}\Delta K)^{2}+V_{i}^{2}}.\label{eq:et2dota}
\end{align}
We see that near the maxima of $\TTT(E)$, with $\Delta K\to0$, the reflection
delay strongly deviates from its $\VI=0$ shape owing to the large negative term
$\dot{\eta}_{2}(K_{n}^{\mathrm{res}})\sim-1/\VI$. This is illustrated in Fig.~\ref{fig4},
where the reflection time delay is shown for the same set of parameters as in
Fig.~\ref{fig3}. As follows from Eq.~(\ref{eq:et2dota}), in the vicinity of the
resonance, the reflection time delay $\TTR(E)$ strongly deviates from $\TTT(E)$
showing a deep minimum, which is the sharper the smaller $\VI$. This instability
is apparently an artifact of the optical-potential-based phenomenology, and it points
to its limitations: Indeed, the divergent time delay at $\VI\to0$ can hardly be
ascribed a physical meaning.

\begin{figure}[b!] 
\centering 
\includegraphics[width=\columnwidth]{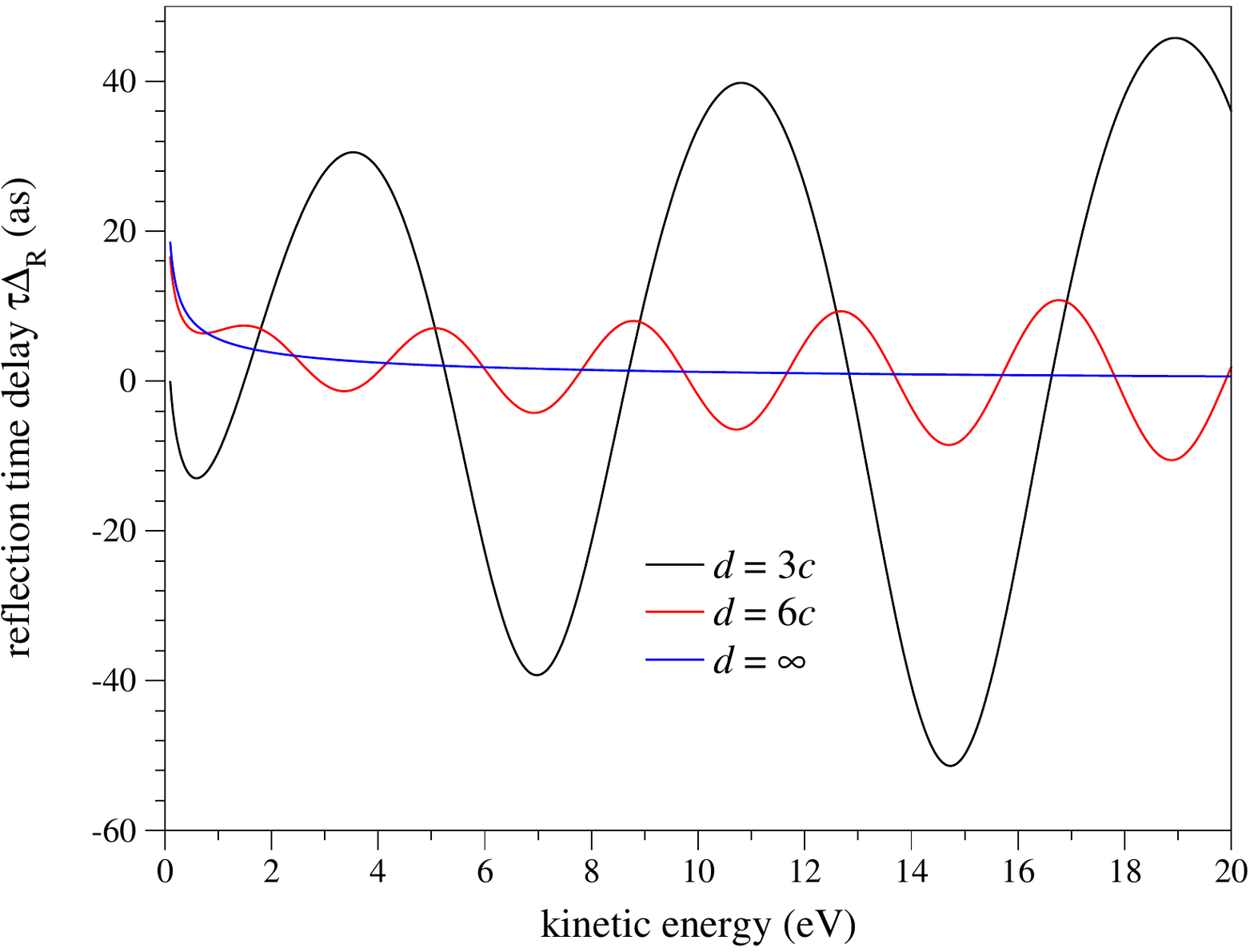}
\caption{Reflection time delay $\TTR$ for the absorbing potential
  value $\VI=3$~eV, and the well width $d=3c$ and $6c$, as well as for
  the semi-infinite well $d=\infty$. }
\label{fig5} 
\end{figure}   
Furthermore, in the limit of semi-infinite crystal, the energy dependent imaginary
part of $K$ caused by the absorbing potential leads to a finite reflection time
delay because the reflection amplitude $r_{\infty}$ in Eq. (\ref{eq:rinfreal})
becomes complex. Figure~\ref{fig5} shows the reflection time delay for a fixed
value of $\VI=3$~eV for different slab widths. As one would expect, in the presence
of the absorbing potential, with the increase of the slab width $d$ the reflection
time delay steadily converges to its value for the semi-infinite crystal.

\section{Conclusion\label{sec:Conclusion}}

The present study demonstrates that the theory of attosecond dynamics
requires a careful analysis of the phenomenology that underlies the
calculations. In particular, we have studied the effect of the absorbing
potential on the phase time (EWS delay) of the transmitted and reflected
wave packet. For the transmitted packet the effect of $\VI$ is to damp out
the oscillations of the transit time. Thereby in the limit $\VI\to\infty$ the
oscillating $\TTT(E)$ curve converges to the intuitively understandable
Landauer-Martin curve.

Contrastingly, for the reflected packet the effect is rather counterintuitive:
as a function of $\VI$, the time delay $\TTR(E)$ shows a discontinuity at its
maxima, with $\TTR(E^{\rm max})\to-\infty$ at $\VI\to0$. Furthermore, at realistic
values of $\VI$ the reflection delay shows appreciably large negative values
(i.e., time advance) over an energy interval of tenths of eV. This behavior is
common to transmission resonances in various 1D systems, and here we have
demonstrated it for a simple system amenable to analytical treatment and
obtained the exact expressions~(\ref{eq:tauWr}) and~(\ref{eq:et2dota}),
leaving no doubts as to the validity of this result.

The large negative delays do not seem to have physical meaning, which raises
some concerns regarding the applicability of the EWS formalism in combination
with the phenomenology of optical potential. In view of the wide application
of both concepts in transport, LEED, and photoemission theories, this calls
for seeking out alternative ways of allowing for the inelastic scattering. This
is quite challenging because, for example, introducing the non-hermiticity by
means of nonreciprocal hopping is limited to the tight-binding representation,
which is not applicable above the vacuum level. Generally, it is not clear how
to reconcile the actual dephasing of the wave packet with the phase-time
formalism. In any case, in using the absorbing-potential approach it is
important to be aware of possible potholes and carefully check the physical
consistency of the results. To our knowledge, the $\VI$-related spurious
features of LEED are here revealed for the first time, and we hope the present
study to draw further attention to the problem.

\begin{acknowledgments}
This work was supported by the Spanish Ministry of Science and Innovation
(MICINN Project No.~PID2022-139230NB-I00) and by the National Academy
of Sciences of Ukraine (Project No.~III-2-22 and III-4-23)).
\end{acknowledgments}

\appendix
\section{Details of derivation \label{sec:Appendix}} 

The condition of the continuity of the function $\psi(z)$ and its derivative
$\partial\psi/\partial z$ at $z=0$ is 
\begin{equation}
\begin{cases}
\hfill 1+r&=A+B,\\
k(1-r)&=K(A-B).
\end{cases}\label{eq:bcz0}
\end{equation}
The continuity at $z=d$ yields 
\begin{equation}
\begin{cases}
\hfill Ae^{ikd}+Be^{-iKd} &=te^{ikd}\\
K\left(Ae^{iKd}-Be^{-iKd}\right)&=kte^{ikd}.
\end{cases}\label{eq:bczb}
\end{equation}
Excluding $t\exp(ikd)$ from the system (\ref{eq:bczb}), we obtain
\[
K\left(Ae^{iKd}-Be^{-iKd}\right)=k\left(Ae^{iKd}+Be^{-iKd}\right),
\]
which gives the relation between the amplitudes $A$ and $B$:
\[
B=Ae^{iKd}\frac{K-k}{K+k}.
\]
Then from the first equation of the system (\ref{eq:bczb}) we express
$A$ via $t\exp(ikd)$. Finally we find the expressions (\ref{eq:A})
and (\ref{eq:B}) for both amplitudes:
\begin{align}
A & =\frac{K+k}{2K}te^{i\left(k-K\right)d},\label{eq:aA}\\
B & =\frac{K-k}{2K}te^{i\left(k+K\right)d}.\label{eq:aB}
\end{align}
We rewrite the system (\ref{eq:bcz0}) as
\begin{equation}
\begin{cases}
1+r=A+B,\\
1-r=\dfrac{K}{k}(A-B).
\end{cases}\label{eq:bcz0-1}
\end{equation}
By substituting Eqs.~(\ref{eq:aA}) and (\ref{eq:aB}) into the
system~(\ref{eq:bcz0-1}) we obtain Eq.~(\ref{eq:t}) from the sum of the
two equations~(\ref{eq:bcz0-1}) and Eq.~(\ref{eq:r}) from their difference. 
Further, we obtain the equations (\ref{eq:A})--(\ref{eq:B}).

The reflection from the half space is considered similarly. The region
$z>d$ is absent, and the solution in the right half-space is $\psi(z)=A\exp(iKz)$.
The only boundary condition is at $z=0$; it is given by Eqs.~(\ref{eq:bcz0})
with $B=0$. This immediately yields $k(1-r_{\infty})=K(1+r_{\infty})$, 
which leads to Eq.~(\ref{eq:rinfreal}).

The boundary conditions~(\ref{eq:bcz0}) and (\ref{eq:bczb}) are the
same for real and complex $U$. Thus, the expressions for the wave
function coefficients (\ref{eq:t})--(\ref{eq:B}) have the same form
in the presence of absorbing potential $-i\VI$. However, the wave
vector inside the slab becomes complex, $K=K_{1}+iK_{2}$, and the dependence 
of the phases of the coefficients on energy becomes more sophisticated:
\begin{widetext}
\begin{flalign}
\sin(Kd) & =\sin(K_{1}d)\cosh(K_{2}d)+i\cos(K_{1}d)\sinh(K_{2}d),&& \nonumber \\
\cos(Kd) & =\cos(K_{1}d)\cosh(K_{2}d)-i\sin(K_{1}d)\sinh(K_{2}d),&& \nonumber
\end{flalign}
etc. The transmission amplitude now has the form (\ref{eq:teikdVi}),
similar to (\ref{eq:teikd}), but the expressions for $p$ and $q$
become formidable 
\begin{flalign}
p & =\cos(K_{1}d)\bigg[\cosh(K_{2}d) 
+\frac{K_{1}}{2}\left(\frac{1}{k}
+\frac{k}{|K|^{2}}\right)\sinh(K_{2}d)\bigg]
+\frac{K_{2}}{2}\left(\frac{1}{k}
-\frac{k}{|K|^{2}}\right)\sin(K_{1}d)\cosh(K_{2}d), && \label{eq:at}\\
q & =\sin(K_{1}d)\bigg[\sinh(K_{2}d)
+\frac{K_{1}}{2}\left(\frac{1}{k}+\frac{k}{|K|^{2}}\right)\cosh(K_{2}d)\bigg]
 -\frac{K_{2}}{2}\left(\frac{1}{k}
 -\frac{k}{|K|^{2}}\right)\cos\left(K_{1}d\right)\sinh\left(K_{2}d\right). &&
\label{eq:bt}
\end{flalign}
For the calculation of the transit time, Eq.~(\ref{eq:etatdot}), we need
the energy derivatives of these terms 
\begin{flalign}
\dot{p} =& -\sin(K_{1}d)d\dot{K}_{1}\bigg[\cosh(K_{2}d) +\frac{K_{1}}{2}\left(\frac{1}{k}
+\frac{k}{|K|^{2}}\right)\sinh(K_{2}d)\bigg] && \nonumber \\
&+\cos(K_{1}d)\bigg\{\sinh(K_{2}d)d\dot{K}_{2} +\frac{\partial}{\partial E}
\left[\frac{K_{1}}{2}\left(\frac{1}{k}+\frac{k}{|K|^{2}}\right)\right]\sinh(K_{2}d)
+\frac{K_{1}}{2}\left(\frac{1}{k}
+\frac{k}{|K|^{2}}\right)\cosh\left(K_{2}d\right)d\dot{K}_{2}\bigg\} && \nonumber \\
&+\frac{\partial}{\partial E}\left[\frac{K_{2}}{2}\left(\frac{1}{k}
-\frac{k}{|K|^{2}}\right)\right]\sin(K_{1}d)\cosh(K_{2}d) && \nonumber \\
&+\frac{K_{2}}{2}\left(\frac{1}{k}-\frac{k}{|K|^{2}}\right)
\big[\cos\left(K_{1}d\right)d\dot{K}_{1}\cosh\left(K_{2}d\right) 
+\sin\left(K_{1}d\right)\sinh\left(K_{2}d\right)d\dot{K}_{2}\big], && \label{eq:atdot}
\end{flalign}
where
\begin{flalign}
\frac{\partial}{\partial E}\left[\frac{K_{1}}{2}\left(\frac{1}{k}
  +\frac{k}{|K|^{2}}\right)\right]
& =\frac{\dot{K}_{1}}{2}\left(\frac{1}{k}+\frac{k}{|K|^{2}}\right) 
+\frac{K_{1}}{2}\bigg[-\frac{1}{k^{3}}+\frac{1}{k|K|^{2}}
-\frac{2k}{|K|^{4}}\left(K_{1}\dot{K}_{1}+K_{2}\dot{K}_{2}\right)\bigg], && \nonumber \\
\frac{\partial}{\partial E}
 \left[\frac{K_{2}}{2}\left(\frac{1}{k}-\frac{k}{|K|^{2}}\right)\right] 
& =\frac{\dot{K}_{2}}{2}\left(\frac{1}{k}-\frac{k}{|K|^{2}}\right) 
 +\frac{K_{2}}{2}\bigg[-\frac{1}{k^{3}}-\frac{1}{k|K|^{2}}+\frac{2k}{|K|^{4}}
 \left(K_{1}\dot{K}_{1}+K_{2}\dot{K}_{2}\right)\bigg], && \nonumber
\end{flalign}
and
\begin{flalign}
\dot{q} & =\cos(K_{1}d)d\dot{K}_{1}\bigg[\sinh(K_{2}d) 
+\frac{K_{1}}{2}\left(\frac{1}{k} +\frac{k}{|K|^{2}}\right)\cosh(K_{2}d)\bigg] && \nonumber \\
& +\sin(K_{1}d)\bigg\{\cosh(K_{2}d)d\dot{K}_{2} 
  +\frac{\partial}{\partial E}\left[\frac{K_{1}}{2}\left(\frac{1}{k}
  +\frac{k}{|K|^{2}}\right)\right]\cosh(K_{2}d) 
  +\frac{K_{1}}{2}\left(\frac{1}{k}+\frac{k}{|K|^{2}}\right)
\sinh\left(K_{2}d\right)d\dot{K}_{2}\bigg\} && \nonumber \\
& -\frac{\partial}{\partial E}\left[\frac{K_{2}}{2}
\left(\frac{1}{k}-\frac{k}{|K|^{2}}\right)\right]\cos\left(K_{1}d\right)\sinh\left(K_{2}d\right)
&& \nonumber \\
& -\frac{K_{2}}{2}\left(\frac{1}{k} -\frac{k}{|K|^{2}}\right)
\big[-\sin(K_{1}d)d\dot{K}_{1}\sinh(K_{2}d) 
+\cos(K_{1}d)\cosh(K_{2}d)d\dot{K}_{2}\big]. && \label{eq:btdot}
\end{flalign}
\end{widetext}

\bibliography{jel}

\begin{thebibliography}{32}%
\makeatletter
\providecommand \@ifxundefined [1]{%
 \@ifx{#1\undefined}
}%
\providecommand \@ifnum [1]{%
 \ifnum #1\expandafter \@firstoftwo
 \else \expandafter \@secondoftwo
 \fi
}%
\providecommand \@ifx [1]{%
 \ifx #1\expandafter \@firstoftwo
 \else \expandafter \@secondoftwo
 \fi
}%
\providecommand \natexlab [1]{#1}%
\providecommand \enquote  [1]{``#1''}%
\providecommand \bibnamefont  [1]{#1}%
\providecommand \bibfnamefont [1]{#1}%
\providecommand \citenamefont [1]{#1}%
\providecommand \href@noop [0]{\@secondoftwo}%
\providecommand \href [0]{\begingroup \@sanitize@url \@href}%
\providecommand \@href[1]{\@@startlink{#1}\@@href}%
\providecommand \@@href[1]{\endgroup#1\@@endlink}%
\providecommand \@sanitize@url [0]{\catcode `\\12\catcode `\$12\catcode
  `\&12\catcode `\#12\catcode `\^12\catcode `\_12\catcode `\%12\relax}%
\providecommand \@@startlink[1]{}%
\providecommand \@@endlink[0]{}%
\providecommand \url  [0]{\begingroup\@sanitize@url \@url }%
\providecommand \@url [1]{\endgroup\@href {#1}{\urlprefix }}%
\providecommand \urlprefix  [0]{URL }%
\providecommand \Eprint [0]{\href }%
\providecommand \doibase [0]{https://doi.org/}%
\providecommand \selectlanguage [0]{\@gobble}%
\providecommand \bibinfo  [0]{\@secondoftwo}%
\providecommand \bibfield  [0]{\@secondoftwo}%
\providecommand \translation [1]{[#1]}%
\providecommand \BibitemOpen [0]{}%
\providecommand \bibitemStop [0]{}%
\providecommand \bibitemNoStop [0]{.\EOS\space}%
\providecommand \EOS [0]{\spacefactor3000\relax}%
\providecommand \BibitemShut  [1]{\csname bibitem#1\endcsname}%
\let\auto@bib@innerbib\@empty
\bibitem [{\citenamefont {Cavalieri}\ \emph {et~al.}(2007)\citenamefont
  {Cavalieri}, \citenamefont {Mueller}, \citenamefont {Uphues}, \citenamefont
  {Yakovlev}, \citenamefont {Baltuska}, \citenamefont {Horvath}, \citenamefont
  {Schmidt}, \citenamefont {Bluemel}, \citenamefont {Holzwarth}, \citenamefont
  {Hendel}, \citenamefont {Drescher}, \citenamefont {Kleineberg}, \citenamefont
  {Echenique}, \citenamefont {Kienberger}, \citenamefont {Krausz},\ and\
  \citenamefont {Heinzmann}}]{Cavalieri2007}%
  \BibitemOpen
  \bibfield  {author} {\bibinfo {author} {\bibfnamefont {A.~L.}\ \bibnamefont
  {Cavalieri}}, \bibinfo {author} {\bibfnamefont {N.}~\bibnamefont {Mueller}},
  \bibinfo {author} {\bibfnamefont {T.}~\bibnamefont {Uphues}}, \bibinfo
  {author} {\bibfnamefont {V.~S.}\ \bibnamefont {Yakovlev}}, \bibinfo {author}
  {\bibfnamefont {A.}~\bibnamefont {Baltuska}}, \bibinfo {author}
  {\bibfnamefont {B.}~\bibnamefont {Horvath}}, \bibinfo {author} {\bibfnamefont
  {B.}~\bibnamefont {Schmidt}}, \bibinfo {author} {\bibfnamefont
  {L.}~\bibnamefont {Bluemel}}, \bibinfo {author} {\bibfnamefont
  {R.}~\bibnamefont {Holzwarth}}, \bibinfo {author} {\bibfnamefont
  {S.}~\bibnamefont {Hendel}}, \bibinfo {author} {\bibfnamefont
  {M.}~\bibnamefont {Drescher}}, \bibinfo {author} {\bibfnamefont
  {U.}~\bibnamefont {Kleineberg}}, \bibinfo {author} {\bibfnamefont {P.~M.}\
  \bibnamefont {Echenique}}, \bibinfo {author} {\bibfnamefont {R.}~\bibnamefont
  {Kienberger}}, \bibinfo {author} {\bibfnamefont {F.}~\bibnamefont {Krausz}},\
  and\ \bibinfo {author} {\bibfnamefont {U.}~\bibnamefont {Heinzmann}},\
  }\bibfield  {title} {\bibinfo {title} {Attosecond spectroscopy in condensed
  matter},\ }\href {https://doi.org/10.1038/nature06229} {\bibfield  {journal}
  {\bibinfo  {journal} {Nature}\ }\textbf {\bibinfo {volume} {449}},\ \bibinfo
  {pages} {1029} (\bibinfo {year} {2007})}\BibitemShut {NoStop}%
\bibitem [{\citenamefont {Schultze}\ \emph {et~al.}(2010)\citenamefont
  {Schultze}, \citenamefont {Fie{\ss}}, \citenamefont {Karpowicz},
  \citenamefont {Gagnon}, \citenamefont {Korbman}, \citenamefont {Hofstetter},
  \citenamefont {Neppl}, \citenamefont {Cavalieri}, \citenamefont {Komninos},
  \citenamefont {Mercouris}, \citenamefont {Nicolaides}, \citenamefont
  {Pazourek}, \citenamefont {Nagele}, \citenamefont {Feist}, \citenamefont
  {Burgd{\"o}rfer}, \citenamefont {Azzeer}, \citenamefont {Ernstorfer},
  \citenamefont {Kienberger}, \citenamefont {Kleineberg}, \citenamefont
  {Goulielmakis}, \citenamefont {Krausz},\ and\ \citenamefont
  {Yakovlev}}]{Schultze10}%
  \BibitemOpen
  \bibfield  {author} {\bibinfo {author} {\bibfnamefont {M.}~\bibnamefont
  {Schultze}}, \bibinfo {author} {\bibfnamefont {M.}~\bibnamefont {Fie{\ss}}},
  \bibinfo {author} {\bibfnamefont {N.}~\bibnamefont {Karpowicz}}, \bibinfo
  {author} {\bibfnamefont {J.}~\bibnamefont {Gagnon}}, \bibinfo {author}
  {\bibfnamefont {M.}~\bibnamefont {Korbman}}, \bibinfo {author} {\bibfnamefont
  {M.}~\bibnamefont {Hofstetter}}, \bibinfo {author} {\bibfnamefont
  {S.}~\bibnamefont {Neppl}}, \bibinfo {author} {\bibfnamefont {A.~L.}\
  \bibnamefont {Cavalieri}}, \bibinfo {author} {\bibfnamefont {Y.}~\bibnamefont
  {Komninos}}, \bibinfo {author} {\bibfnamefont {T.}~\bibnamefont {Mercouris}},
  \bibinfo {author} {\bibfnamefont {C.~A.}\ \bibnamefont {Nicolaides}},
  \bibinfo {author} {\bibfnamefont {R.}~\bibnamefont {Pazourek}}, \bibinfo
  {author} {\bibfnamefont {S.}~\bibnamefont {Nagele}}, \bibinfo {author}
  {\bibfnamefont {J.}~\bibnamefont {Feist}}, \bibinfo {author} {\bibfnamefont
  {J.}~\bibnamefont {Burgd{\"o}rfer}}, \bibinfo {author} {\bibfnamefont
  {A.~M.}\ \bibnamefont {Azzeer}}, \bibinfo {author} {\bibfnamefont
  {R.}~\bibnamefont {Ernstorfer}}, \bibinfo {author} {\bibfnamefont
  {R.}~\bibnamefont {Kienberger}}, \bibinfo {author} {\bibfnamefont
  {U.}~\bibnamefont {Kleineberg}}, \bibinfo {author} {\bibfnamefont
  {E.}~\bibnamefont {Goulielmakis}}, \bibinfo {author} {\bibfnamefont
  {F.}~\bibnamefont {Krausz}},\ and\ \bibinfo {author} {\bibfnamefont {V.~S.}\
  \bibnamefont {Yakovlev}},\ }\bibfield  {title} {\bibinfo {title} {Delay in
  photoemission},\ }\href {https://doi.org/10.1126/science.1189401} {\bibfield
  {journal} {\bibinfo  {journal} {Science}\ }\textbf {\bibinfo {volume}
  {328}},\ \bibinfo {pages} {1658} (\bibinfo {year} {2010})}\BibitemShut
  {NoStop}%
\bibitem [{\citenamefont {Neppl}\ \emph {et~al.}(2012)\citenamefont {Neppl},
  \citenamefont {Ernstorfer}, \citenamefont {Bothschafter}, \citenamefont
  {Cavalieri}, \citenamefont {Menzel}, \citenamefont {Barth}, \citenamefont
  {Krausz}, \citenamefont {Kienberger},\ and\ \citenamefont
  {Feulner}}]{Neppl2012}%
  \BibitemOpen
  \bibfield  {author} {\bibinfo {author} {\bibfnamefont {S.}~\bibnamefont
  {Neppl}}, \bibinfo {author} {\bibfnamefont {R.}~\bibnamefont {Ernstorfer}},
  \bibinfo {author} {\bibfnamefont {E.~M.}\ \bibnamefont {Bothschafter}},
  \bibinfo {author} {\bibfnamefont {A.~L.}\ \bibnamefont {Cavalieri}}, \bibinfo
  {author} {\bibfnamefont {D.}~\bibnamefont {Menzel}}, \bibinfo {author}
  {\bibfnamefont {J.~V.}\ \bibnamefont {Barth}}, \bibinfo {author}
  {\bibfnamefont {F.}~\bibnamefont {Krausz}}, \bibinfo {author} {\bibfnamefont
  {R.}~\bibnamefont {Kienberger}},\ and\ \bibinfo {author} {\bibfnamefont
  {P.}~\bibnamefont {Feulner}},\ }\bibfield  {title} {\bibinfo {title}
  {Attosecond time-resolved photoemission from core and valence states of
  magnesium},\ }\href {https://doi.org/10.1103/PhysRevLett.109.087401}
  {\bibfield  {journal} {\bibinfo  {journal} {Phys. Rev. Lett.}\ }\textbf
  {\bibinfo {volume} {109}},\ \bibinfo {pages} {087401} (\bibinfo {year}
  {2012})}\BibitemShut {NoStop}%
\bibitem [{\citenamefont {Neppl}\ \emph {et~al.}(2015)\citenamefont {Neppl},
  \citenamefont {Ernstorfer}, \citenamefont {Cavalieri}, \citenamefont
  {Lemell}, \citenamefont {Wachter}, \citenamefont {Magerl}, \citenamefont
  {Bothschafter}, \citenamefont {Jobst}, \citenamefont {Hofstetter},
  \citenamefont {Kleineberg}, \citenamefont {Barth}, \citenamefont {Menzel},
  \citenamefont {Burgd{\"o}rfer}, \citenamefont {Feulner}, \citenamefont
  {Krausz},\ and\ \citenamefont {Kienberger}}]{Neppl15}%
  \BibitemOpen
  \bibfield  {author} {\bibinfo {author} {\bibfnamefont {S.}~\bibnamefont
  {Neppl}}, \bibinfo {author} {\bibfnamefont {R.}~\bibnamefont {Ernstorfer}},
  \bibinfo {author} {\bibfnamefont {A.~L.}\ \bibnamefont {Cavalieri}}, \bibinfo
  {author} {\bibfnamefont {C.}~\bibnamefont {Lemell}}, \bibinfo {author}
  {\bibfnamefont {G.}~\bibnamefont {Wachter}}, \bibinfo {author} {\bibfnamefont
  {E.}~\bibnamefont {Magerl}}, \bibinfo {author} {\bibfnamefont {E.~M.}\
  \bibnamefont {Bothschafter}}, \bibinfo {author} {\bibfnamefont
  {M.}~\bibnamefont {Jobst}}, \bibinfo {author} {\bibfnamefont
  {M.}~\bibnamefont {Hofstetter}}, \bibinfo {author} {\bibfnamefont
  {U.}~\bibnamefont {Kleineberg}}, \bibinfo {author} {\bibfnamefont {J.~V.}\
  \bibnamefont {Barth}}, \bibinfo {author} {\bibfnamefont {D.}~\bibnamefont
  {Menzel}}, \bibinfo {author} {\bibfnamefont {J.}~\bibnamefont
  {Burgd{\"o}rfer}}, \bibinfo {author} {\bibfnamefont {P.}~\bibnamefont
  {Feulner}}, \bibinfo {author} {\bibfnamefont {F.}~\bibnamefont {Krausz}},\
  and\ \bibinfo {author} {\bibfnamefont {R.}~\bibnamefont {Kienberger}},\
  }\bibfield  {title} {\bibinfo {title} {Direct observation of electron
  propagation and dielectric screening on the atomic length scale},\ }\href
  {https://doi.org/10.1038/nature14094} {\bibfield  {journal} {\bibinfo
  {journal} {Nature}\ }\textbf {\bibinfo {volume} {517}},\ \bibinfo {pages}
  {342} (\bibinfo {year} {2015})}\BibitemShut {NoStop}%
\bibitem [{\citenamefont {Okell}\ \emph {et~al.}(2015)\citenamefont {Okell},
  \citenamefont {Witting}, \citenamefont {Fabris}, \citenamefont {Arrell},
  \citenamefont {Hengster}, \citenamefont {Ibrahimkutty}, \citenamefont
  {Seiler}, \citenamefont {Barthelmess}, \citenamefont {Stankov}, \citenamefont
  {Lei}, \citenamefont {Sonnefraud}, \citenamefont {Rahmani}, \citenamefont
  {Uphues}, \citenamefont {Maier}, \citenamefont {Marangos},\ and\
  \citenamefont {Tisch}}]{Okell15}%
  \BibitemOpen
  \bibfield  {author} {\bibinfo {author} {\bibfnamefont {W.~A.}\ \bibnamefont
  {Okell}}, \bibinfo {author} {\bibfnamefont {T.}~\bibnamefont {Witting}},
  \bibinfo {author} {\bibfnamefont {D.}~\bibnamefont {Fabris}}, \bibinfo
  {author} {\bibfnamefont {C.~A.}\ \bibnamefont {Arrell}}, \bibinfo {author}
  {\bibfnamefont {J.}~\bibnamefont {Hengster}}, \bibinfo {author}
  {\bibfnamefont {S.}~\bibnamefont {Ibrahimkutty}}, \bibinfo {author}
  {\bibfnamefont {A.}~\bibnamefont {Seiler}}, \bibinfo {author} {\bibfnamefont
  {M.}~\bibnamefont {Barthelmess}}, \bibinfo {author} {\bibfnamefont
  {S.}~\bibnamefont {Stankov}}, \bibinfo {author} {\bibfnamefont {D.~Y.}\
  \bibnamefont {Lei}}, \bibinfo {author} {\bibfnamefont {Y.}~\bibnamefont
  {Sonnefraud}}, \bibinfo {author} {\bibfnamefont {M.}~\bibnamefont {Rahmani}},
  \bibinfo {author} {\bibfnamefont {T.}~\bibnamefont {Uphues}}, \bibinfo
  {author} {\bibfnamefont {S.~A.}\ \bibnamefont {Maier}}, \bibinfo {author}
  {\bibfnamefont {J.~P.}\ \bibnamefont {Marangos}},\ and\ \bibinfo {author}
  {\bibfnamefont {J.~W.~G.}\ \bibnamefont {Tisch}},\ }\bibfield  {title}
  {\bibinfo {title} {Temporal broadening of attosecond photoelectron
  wavepackets from solid surfaces},\ }\href
  {https://doi.org/10.1364/OPTICA.2.000383} {\bibfield  {journal} {\bibinfo
  {journal} {Optica}\ }\textbf {\bibinfo {volume} {2}},\ \bibinfo {pages} {383}
  (\bibinfo {year} {2015})}\BibitemShut {NoStop}%
\bibitem [{\citenamefont {Siek}\ \emph {et~al.}(2017)\citenamefont {Siek},
  \citenamefont {Neb}, \citenamefont {Bartz}, \citenamefont {Hensen},
  \citenamefont {Str\"{u}ber}, \citenamefont {Fiechter}, \citenamefont
  {Torrent-Sucarrat}, \citenamefont {Silkin}, \citenamefont {Krasovskii},
  \citenamefont {Kabachnik}, \citenamefont {Fritzsche}, \citenamefont {{n}o},
  \citenamefont {Echenique}, \citenamefont {Kazansky}, \citenamefont
  {M\"{u}ller}, \citenamefont {Pfeiffer},\ and\ \citenamefont
  {Heinzmann}}]{Siek2017}%
  \BibitemOpen
  \bibfield  {author} {\bibinfo {author} {\bibfnamefont {F.}~\bibnamefont
  {Siek}}, \bibinfo {author} {\bibfnamefont {S.}~\bibnamefont {Neb}}, \bibinfo
  {author} {\bibfnamefont {P.}~\bibnamefont {Bartz}}, \bibinfo {author}
  {\bibfnamefont {M.}~\bibnamefont {Hensen}}, \bibinfo {author} {\bibfnamefont
  {C.}~\bibnamefont {Str\"{u}ber}}, \bibinfo {author} {\bibfnamefont
  {S.}~\bibnamefont {Fiechter}}, \bibinfo {author} {\bibfnamefont
  {M.}~\bibnamefont {Torrent-Sucarrat}}, \bibinfo {author} {\bibfnamefont
  {V.~M.}\ \bibnamefont {Silkin}}, \bibinfo {author} {\bibfnamefont {E.~E.}\
  \bibnamefont {Krasovskii}}, \bibinfo {author} {\bibfnamefont
  {N.}~\bibnamefont {Kabachnik}}, \bibinfo {author} {\bibfnamefont
  {S.}~\bibnamefont {Fritzsche}}, \bibinfo {author} {\bibfnamefont {R.~D.~M.}\
  \bibnamefont {{n}o}}, \bibinfo {author} {\bibfnamefont {P.~M.}\ \bibnamefont
  {Echenique}}, \bibinfo {author} {\bibfnamefont {A.~K.}\ \bibnamefont
  {Kazansky}}, \bibinfo {author} {\bibfnamefont {N.}~\bibnamefont
  {M\"{u}ller}}, \bibinfo {author} {\bibfnamefont {W.}~\bibnamefont
  {Pfeiffer}},\ and\ \bibinfo {author} {\bibfnamefont {U.}~\bibnamefont
  {Heinzmann}},\ }\bibfield  {title} {\bibinfo {title} {Angular
  momentum-induced delays in solid-state photoemission enhanced by intra-atomic
  interactions},\ }\href {https://doi.org/10.1126/science.aam9598} {\bibfield
  {journal} {\bibinfo  {journal} {Science}\ }\textbf {\bibinfo {volume}
  {357}},\ \bibinfo {pages} {1274} (\bibinfo {year} {2017})}\BibitemShut
  {NoStop}%
\bibitem [{\citenamefont {Ossiander}\ \emph {et~al.}(2018)\citenamefont
  {Ossiander}, \citenamefont {Riemensberger}, \citenamefont {Neppl},
  \citenamefont {Mittermair}, \citenamefont {Sch\"affer}, \citenamefont
  {Duensing}, \citenamefont {Wagner}, \citenamefont {Heider}, \citenamefont
  {Wurzer}, \citenamefont {Gerl}, \citenamefont {Schnitzenbaumer},
  \citenamefont {Barth}, \citenamefont {Libisch}, \citenamefont {Lemell},
  \citenamefont {Burgd\"orfer}, \citenamefont {Feulner},\ and\ \citenamefont
  {Kienberger}}]{Ossiander2018}%
  \BibitemOpen
  \bibfield  {author} {\bibinfo {author} {\bibfnamefont {M.}~\bibnamefont
  {Ossiander}}, \bibinfo {author} {\bibfnamefont {J.}~\bibnamefont
  {Riemensberger}}, \bibinfo {author} {\bibfnamefont {S.}~\bibnamefont
  {Neppl}}, \bibinfo {author} {\bibfnamefont {M.}~\bibnamefont {Mittermair}},
  \bibinfo {author} {\bibfnamefont {M.}~\bibnamefont {Sch\"affer}}, \bibinfo
  {author} {\bibfnamefont {A.}~\bibnamefont {Duensing}}, \bibinfo {author}
  {\bibfnamefont {M.~S.}\ \bibnamefont {Wagner}}, \bibinfo {author}
  {\bibfnamefont {R.}~\bibnamefont {Heider}}, \bibinfo {author} {\bibfnamefont
  {M.}~\bibnamefont {Wurzer}}, \bibinfo {author} {\bibfnamefont
  {M.}~\bibnamefont {Gerl}}, \bibinfo {author} {\bibfnamefont {M.}~\bibnamefont
  {Schnitzenbaumer}}, \bibinfo {author} {\bibfnamefont {J.~V.}\ \bibnamefont
  {Barth}}, \bibinfo {author} {\bibfnamefont {F.}~\bibnamefont {Libisch}},
  \bibinfo {author} {\bibfnamefont {C.}~\bibnamefont {Lemell}}, \bibinfo
  {author} {\bibfnamefont {J.}~\bibnamefont {Burgd\"orfer}}, \bibinfo {author}
  {\bibfnamefont {P.}~\bibnamefont {Feulner}},\ and\ \bibinfo {author}
  {\bibfnamefont {R.}~\bibnamefont {Kienberger}},\ }\bibfield  {title}
  {\bibinfo {title} {Absolute timing of the photoelectric effect},\ }\href
  {https://doi.org/10.1038/s41586-018-0503-6} {\bibfield  {journal} {\bibinfo
  {journal} {Nature}\ }\textbf {\bibinfo {volume} {561}},\ \bibinfo {pages}
  {374} (\bibinfo {year} {2018})}\BibitemShut {NoStop}%
\bibitem [{\citenamefont {Riemensberger}\ \emph {et~al.}(2019)\citenamefont
  {Riemensberger}, \citenamefont {Neppl}, \citenamefont {Potamianos},
  \citenamefont {Sch\"affer}, \citenamefont {Schnitzenbaumer}, \citenamefont
  {Ossiander}, \citenamefont {Schr\"oder}, \citenamefont {Guggenmos},
  \citenamefont {Kleineberg}, \citenamefont {Menzel}, \citenamefont
  {Allegretti}, \citenamefont {Barth}, \citenamefont {Kienberger},
  \citenamefont {Feulner}, \citenamefont {Borisov}, \citenamefont {Echenique},\
  and\ \citenamefont {Kazansky}}]{Riemensberger2019}%
  \BibitemOpen
  \bibfield  {author} {\bibinfo {author} {\bibfnamefont {J.}~\bibnamefont
  {Riemensberger}}, \bibinfo {author} {\bibfnamefont {S.}~\bibnamefont
  {Neppl}}, \bibinfo {author} {\bibfnamefont {D.}~\bibnamefont {Potamianos}},
  \bibinfo {author} {\bibfnamefont {M.}~\bibnamefont {Sch\"affer}}, \bibinfo
  {author} {\bibfnamefont {M.}~\bibnamefont {Schnitzenbaumer}}, \bibinfo
  {author} {\bibfnamefont {M.}~\bibnamefont {Ossiander}}, \bibinfo {author}
  {\bibfnamefont {C.}~\bibnamefont {Schr\"oder}}, \bibinfo {author}
  {\bibfnamefont {A.}~\bibnamefont {Guggenmos}}, \bibinfo {author}
  {\bibfnamefont {U.}~\bibnamefont {Kleineberg}}, \bibinfo {author}
  {\bibfnamefont {D.}~\bibnamefont {Menzel}}, \bibinfo {author} {\bibfnamefont
  {F.}~\bibnamefont {Allegretti}}, \bibinfo {author} {\bibfnamefont {J.~V.}\
  \bibnamefont {Barth}}, \bibinfo {author} {\bibfnamefont {R.}~\bibnamefont
  {Kienberger}}, \bibinfo {author} {\bibfnamefont {P.}~\bibnamefont {Feulner}},
  \bibinfo {author} {\bibfnamefont {A.~G.}\ \bibnamefont {Borisov}}, \bibinfo
  {author} {\bibfnamefont {P.~M.}\ \bibnamefont {Echenique}},\ and\ \bibinfo
  {author} {\bibfnamefont {A.~K.}\ \bibnamefont {Kazansky}},\ }\bibfield
  {title} {\bibinfo {title} {Attosecond dynamics of $sp$-band
  photoexcitation},\ }\href {https://doi.org/10.1103/PhysRevLett.123.176801}
  {\bibfield  {journal} {\bibinfo  {journal} {Phys. Rev. Lett.}\ }\textbf
  {\bibinfo {volume} {123}},\ \bibinfo {pages} {176801} (\bibinfo {year}
  {2019})}\BibitemShut {NoStop}%
\bibitem [{\citenamefont {Locher}\ \emph {et~al.}(2015)\citenamefont {Locher},
  \citenamefont {Castiglioni}, \citenamefont {Lucchini}, \citenamefont {Greif},
  \citenamefont {Gallmann}, \citenamefont {Osterwalder}, \citenamefont
  {Hengsberger},\ and\ \citenamefont {Keller}}]{Locher:15}%
  \BibitemOpen
  \bibfield  {author} {\bibinfo {author} {\bibfnamefont {R.}~\bibnamefont
  {Locher}}, \bibinfo {author} {\bibfnamefont {L.}~\bibnamefont {Castiglioni}},
  \bibinfo {author} {\bibfnamefont {M.}~\bibnamefont {Lucchini}}, \bibinfo
  {author} {\bibfnamefont {M.}~\bibnamefont {Greif}}, \bibinfo {author}
  {\bibfnamefont {L.}~\bibnamefont {Gallmann}}, \bibinfo {author}
  {\bibfnamefont {J.}~\bibnamefont {Osterwalder}}, \bibinfo {author}
  {\bibfnamefont {M.}~\bibnamefont {Hengsberger}},\ and\ \bibinfo {author}
  {\bibfnamefont {U.}~\bibnamefont {Keller}},\ }\bibfield  {title} {\bibinfo
  {title} {Energy-dependent photoemission delays from noble metal surfaces by
  attosecond interferometry},\ }\href {https://doi.org/10.1364/OPTICA.2.000405}
  {\bibfield  {journal} {\bibinfo  {journal} {Optica}\ }\textbf {\bibinfo
  {volume} {2}},\ \bibinfo {pages} {405} (\bibinfo {year} {2015})}\BibitemShut
  {NoStop}%
\bibitem [{\citenamefont {Tao}\ \emph {et~al.}(2016)\citenamefont {Tao},
  \citenamefont {Chen}, \citenamefont {Szilv{\'a}si}, \citenamefont {Keller},
  \citenamefont {Mavrikakis}, \citenamefont {Kapteyn},\ and\ \citenamefont
  {Murnane}}]{Tao2016}%
  \BibitemOpen
  \bibfield  {author} {\bibinfo {author} {\bibfnamefont {Z.}~\bibnamefont
  {Tao}}, \bibinfo {author} {\bibfnamefont {C.}~\bibnamefont {Chen}}, \bibinfo
  {author} {\bibfnamefont {T.}~\bibnamefont {Szilv{\'a}si}}, \bibinfo {author}
  {\bibfnamefont {M.}~\bibnamefont {Keller}}, \bibinfo {author} {\bibfnamefont
  {M.}~\bibnamefont {Mavrikakis}}, \bibinfo {author} {\bibfnamefont
  {H.}~\bibnamefont {Kapteyn}},\ and\ \bibinfo {author} {\bibfnamefont
  {M.}~\bibnamefont {Murnane}},\ }\bibfield  {title} {\bibinfo {title} {Direct
  time-domain observation of attosecond final-state lifetimes in photoemission
  from solids},\ }\href {https://doi.org/10.1126/science.aaf6793} {\bibfield
  {journal} {\bibinfo  {journal} {Science}\ }\textbf {\bibinfo {volume}
  {353}},\ \bibinfo {pages} {62} (\bibinfo {year} {2016})}\BibitemShut
  {NoStop}%
\bibitem [{\citenamefont {Bohm}(1951)}]{Bohm1951}%
  \BibitemOpen
  \bibfield  {author} {\bibinfo {author} {\bibfnamefont {D.}~\bibnamefont
  {Bohm}},\ }\href@noop {} {\emph {\bibinfo {title} {Quantum Theory}}}\
  (\bibinfo  {publisher} {Prentice-Hall, New York},\ \bibinfo {year}
  {1951})\BibitemShut {NoStop}%
\bibitem [{\citenamefont {Wigner}(1955)}]{Wigner55}%
  \BibitemOpen
  \bibfield  {author} {\bibinfo {author} {\bibfnamefont {E.~P.}\ \bibnamefont
  {Wigner}},\ }\bibfield  {title} {\bibinfo {title} {Lower limit for the energy
  derivative of the scattering phase shift},\ }\href
  {https://doi.org/10.1103/PhysRev.98.145} {\bibfield  {journal} {\bibinfo
  {journal} {Phys. Rev.}\ }\textbf {\bibinfo {volume} {98}},\ \bibinfo {pages}
  {145} (\bibinfo {year} {1955})}\BibitemShut {NoStop}%
\bibitem [{\citenamefont {Smith}(1960)}]{Smith60}%
  \BibitemOpen
  \bibfield  {author} {\bibinfo {author} {\bibfnamefont {F.~T.}\ \bibnamefont
  {Smith}},\ }\bibfield  {title} {\bibinfo {title} {Lifetime matrix in
  collision theory},\ }\href {https://doi.org/10.1103/PhysRev.118.349}
  {\bibfield  {journal} {\bibinfo  {journal} {Phys. Rev.}\ }\textbf {\bibinfo
  {volume} {118}},\ \bibinfo {pages} {349} (\bibinfo {year}
  {1960})}\BibitemShut {NoStop}%
\bibitem [{\citenamefont {Hauge}\ and\ \citenamefont
  {St\o{}vneng}(1989)}]{Hauge1989}%
  \BibitemOpen
  \bibfield  {author} {\bibinfo {author} {\bibfnamefont {E.~H.}\ \bibnamefont
  {Hauge}}\ and\ \bibinfo {author} {\bibfnamefont {J.~A.}\ \bibnamefont
  {St\o{}vneng}},\ }\bibfield  {title} {\bibinfo {title} {Tunneling times: a
  critical review},\ }\href {https://doi.org/10.1103/RevModPhys.61.917}
  {\bibfield  {journal} {\bibinfo  {journal} {Rev. Mod. Phys.}\ }\textbf
  {\bibinfo {volume} {61}},\ \bibinfo {pages} {917} (\bibinfo {year}
  {1989})}\BibitemShut {NoStop}%
\bibitem [{\citenamefont {Landauer}\ and\ \citenamefont
  {Martin}(1994)}]{Landauer1994}%
  \BibitemOpen
  \bibfield  {author} {\bibinfo {author} {\bibfnamefont {R.}~\bibnamefont
  {Landauer}}\ and\ \bibinfo {author} {\bibfnamefont {T.}~\bibnamefont
  {Martin}},\ }\bibfield  {title} {\bibinfo {title} {Barrier interaction time
  in tunneling},\ }\href {https://doi.org/10.1103/RevModPhys.66.217} {\bibfield
   {journal} {\bibinfo  {journal} {Rev. Mod. Phys.}\ }\textbf {\bibinfo
  {volume} {66}},\ \bibinfo {pages} {217} (\bibinfo {year} {1994})}\BibitemShut
  {NoStop}%
\bibitem [{\citenamefont {Winful}(2006)}]{Winful2006}%
  \BibitemOpen
  \bibfield  {author} {\bibinfo {author} {\bibfnamefont {H.~G.}\ \bibnamefont
  {Winful}},\ }\bibfield  {title} {\bibinfo {title} {Tunneling time, the
  \uppercase{H}artman effect, and superluminality: A proposed resolution of an
  old paradox},\ }\href
  {https://doi.org/http://dx.doi.org/10.1016/j.physrep.2006.09.002} {\bibfield
  {journal} {\bibinfo  {journal} {Physics Reports}\ }\textbf {\bibinfo {volume}
  {436}},\ \bibinfo {pages} {1 } (\bibinfo {year} {2006})}\BibitemShut
  {NoStop}%
\bibitem [{\citenamefont {Field}(2022)}]{Field2022}%
  \BibitemOpen
  \bibfield  {author} {\bibinfo {author} {\bibfnamefont {G.~E.}\ \bibnamefont
  {Field}},\ }\bibfield  {title} {\bibinfo {title} {On the status of quantum
  tunnelling time},\ }\href {https://doi.org/10.1007/s13194-022-00483-9}
  {\bibfield  {journal} {\bibinfo  {journal} {European Journal for Philosophy
  of Science}\ }\textbf {\bibinfo {volume} {12}},\ \bibinfo {pages} {57}
  (\bibinfo {year} {2022})}\BibitemShut {NoStop}%
\bibitem [{\citenamefont {Feibelman}\ and\ \citenamefont
  {Eastman}(1974)}]{FeibelmanEastman_1974}%
  \BibitemOpen
  \bibfield  {author} {\bibinfo {author} {\bibfnamefont {P.~J.}\ \bibnamefont
  {Feibelman}}\ and\ \bibinfo {author} {\bibfnamefont {D.~E.}\ \bibnamefont
  {Eastman}},\ }\bibfield  {title} {\bibinfo {title} {Photoemission
  spectroscopy--correspondence between quantum theory and experimental
  phenomenology},\ }\href {https://doi.org/10.1103/PhysRevB.10.4932} {\bibfield
   {journal} {\bibinfo  {journal} {Phys. Rev. B}\ }\textbf {\bibinfo {volume}
  {10}},\ \bibinfo {pages} {4932} (\bibinfo {year} {1974})}\BibitemShut
  {NoStop}%
\bibitem [{\citenamefont {Slater}(1937)}]{Slater37}%
  \BibitemOpen
  \bibfield  {author} {\bibinfo {author} {\bibfnamefont {J.~C.}\ \bibnamefont
  {Slater}},\ }\bibfield  {title} {\bibinfo {title} {Damped electron waves in
  crystals},\ }\href {https://doi.org/10.1103/PhysRev.51.840} {\bibfield
  {journal} {\bibinfo  {journal} {Phys. Rev.}\ }\textbf {\bibinfo {volume}
  {51}},\ \bibinfo {pages} {840} (\bibinfo {year} {1937})}\BibitemShut
  {NoStop}%
\bibitem [{\citenamefont {Longhi}(2022)}]{Longhi2022}%
  \BibitemOpen
  \bibfield  {author} {\bibinfo {author} {\bibfnamefont {S.}~\bibnamefont
  {Longhi}},\ }\bibfield  {title} {\bibinfo {title} {{Non-Hermitian Hartman
  Effect}},\ }\href {https://doi.org/https://doi.org/10.1002/andp.202200250}
  {\bibfield  {journal} {\bibinfo  {journal} {Annalen der Physik}\ }\textbf
  {\bibinfo {volume} {534}},\ \bibinfo {pages} {2200250} (\bibinfo {year}
  {2022})}\BibitemShut {NoStop}%
\bibitem [{\citenamefont {Jepsen}\ \emph {et~al.}(1982)\citenamefont {Jepsen},
  \citenamefont {Himpsel},\ and\ \citenamefont {Eastman}}]{Jepsen1982}%
  \BibitemOpen
  \bibfield  {author} {\bibinfo {author} {\bibfnamefont {D.~W.}\ \bibnamefont
  {Jepsen}}, \bibinfo {author} {\bibfnamefont {F.~J.}\ \bibnamefont
  {Himpsel}},\ and\ \bibinfo {author} {\bibfnamefont {D.~E.}\ \bibnamefont
  {Eastman}},\ }\bibfield  {title} {\bibinfo {title} {Single-step-model
  analysis of angle-resolved photoemission from \uppercase{N}i(110) and
  \uppercase{C}u(100)},\ }\href {https://doi.org/10.1103/PhysRevB.26.4039}
  {\bibfield  {journal} {\bibinfo  {journal} {Phys. Rev. B}\ }\textbf {\bibinfo
  {volume} {26}},\ \bibinfo {pages} {4039} (\bibinfo {year}
  {1982})}\BibitemShut {NoStop}%
\bibitem [{\citenamefont {Krasovskii}\ \emph {et~al.}(2016)\citenamefont
  {Krasovskii}, \citenamefont {Friedrich}, \citenamefont {Schattke},\ and\
  \citenamefont {Echenique}}]{Krasovskii2016}%
  \BibitemOpen
  \bibfield  {author} {\bibinfo {author} {\bibfnamefont {E.~E.}\ \bibnamefont
  {Krasovskii}}, \bibinfo {author} {\bibfnamefont {C.}~\bibnamefont
  {Friedrich}}, \bibinfo {author} {\bibfnamefont {W.}~\bibnamefont
  {Schattke}},\ and\ \bibinfo {author} {\bibfnamefont {P.~M.}\ \bibnamefont
  {Echenique}},\ }\bibfield  {title} {\bibinfo {title} {{Rapid propagation of a
  Bloch wave packet excited by a femtosecond ultraviolet pulse}},\ }\href
  {https://doi.org/10.1103/PhysRevB.94.195434} {\bibfield  {journal} {\bibinfo
  {journal} {Phys. Rev. B}\ }\textbf {\bibinfo {volume} {94}},\ \bibinfo
  {pages} {195434} (\bibinfo {year} {2016})}\BibitemShut {NoStop}%
\bibitem [{\citenamefont {Pendry}(1976)}]{Pendry76}%
  \BibitemOpen
  \bibfield  {author} {\bibinfo {author} {\bibfnamefont {J.}~\bibnamefont
  {Pendry}},\ }\bibfield  {title} {\bibinfo {title} {Theory of photoemission},\
  }\href {https://doi.org/http://dx.doi.org/10.1016/0039-6028(76)90355-1}
  {\bibfield  {journal} {\bibinfo  {journal} {Surface Science}\ }\textbf
  {\bibinfo {volume} {57}},\ \bibinfo {pages} {679 } (\bibinfo {year}
  {1976})}\BibitemShut {NoStop}%
\bibitem [{\citenamefont {Braun}(1996)}]{Braun96}%
  \BibitemOpen
  \bibfield  {author} {\bibinfo {author} {\bibfnamefont {J.}~\bibnamefont
  {Braun}},\ }\bibfield  {title} {\bibinfo {title} {The theory of
  angle-resolved ultraviolet photoemission and its applications to ordered
  materials},\ }\href {http://stacks.iop.org/0034-4885/59/i=10/a=002}
  {\bibfield  {journal} {\bibinfo  {journal} {Reports on Progress in Physics}\
  }\textbf {\bibinfo {volume} {59}},\ \bibinfo {pages} {1267} (\bibinfo {year}
  {1996})}\BibitemShut {NoStop}%
\bibitem [{\citenamefont {Krasovskii}\ \emph {et~al.}(2002)\citenamefont
  {Krasovskii}, \citenamefont {Schattke}, \citenamefont {Strocov},\ and\
  \citenamefont {Claessen}}]{Krasovskii_2002}%
  \BibitemOpen
  \bibfield  {author} {\bibinfo {author} {\bibfnamefont {E.~E.}\ \bibnamefont
  {Krasovskii}}, \bibinfo {author} {\bibfnamefont {W.}~\bibnamefont
  {Schattke}}, \bibinfo {author} {\bibfnamefont {V.~N.}\ \bibnamefont
  {Strocov}},\ and\ \bibinfo {author} {\bibfnamefont {R.}~\bibnamefont
  {Claessen}},\ }\bibfield  {title} {\bibinfo {title} {{Unoccupied band
  structure of NbSe$_2$ by very low-energy electron diffraction: Experiment and
  theory}},\ }\href {https://doi.org/10.1103/PhysRevB.66.235403} {\bibfield
  {journal} {\bibinfo  {journal} {Phys. Rev. B}\ }\textbf {\bibinfo {volume}
  {66}},\ \bibinfo {pages} {235403} (\bibinfo {year} {2002})}\BibitemShut
  {NoStop}%
\bibitem [{\citenamefont {Delgado}\ \emph {et~al.}(2004)\citenamefont
  {Delgado}, \citenamefont {Muga},\ and\ \citenamefont
  {Ruschhaupt}}]{Delgado2004}%
  \BibitemOpen
  \bibfield  {author} {\bibinfo {author} {\bibfnamefont {F.}~\bibnamefont
  {Delgado}}, \bibinfo {author} {\bibfnamefont {J.~G.}\ \bibnamefont {Muga}},\
  and\ \bibinfo {author} {\bibfnamefont {A.}~\bibnamefont {Ruschhaupt}},\
  }\bibfield  {title} {\bibinfo {title} {Ultrafast propagation of schr\"odinger
  waves in absorbing media},\ }\href
  {https://doi.org/10.1103/PhysRevA.69.022106} {\bibfield  {journal} {\bibinfo
  {journal} {Phys. Rev. A}\ }\textbf {\bibinfo {volume} {69}},\ \bibinfo
  {pages} {022106} (\bibinfo {year} {2004})}\BibitemShut {NoStop}%
\bibitem [{\citenamefont {Falck}\ and\ \citenamefont
  {Hauge}(1988)}]{Falck1988}%
  \BibitemOpen
  \bibfield  {author} {\bibinfo {author} {\bibfnamefont {J.~P.}\ \bibnamefont
  {Falck}}\ and\ \bibinfo {author} {\bibfnamefont {E.~H.}\ \bibnamefont
  {Hauge}},\ }\bibfield  {title} {\bibinfo {title} {Larmor clock reexamined},\
  }\href {https://doi.org/10.1103/PhysRevB.38.3287} {\bibfield  {journal}
  {\bibinfo  {journal} {Phys. Rev. B}\ }\textbf {\bibinfo {volume} {38}},\
  \bibinfo {pages} {3287} (\bibinfo {year} {1988})}\BibitemShut {NoStop}%
\bibitem [{\citenamefont {Hartman}(1962)}]{Hartman1962}%
  \BibitemOpen
  \bibfield  {author} {\bibinfo {author} {\bibfnamefont {T.~E.}\ \bibnamefont
  {Hartman}},\ }\bibfield  {title} {\bibinfo {title} {Tunneling of a wave
  packet},\ }\href {https://doi.org/http://dx.doi.org/10.1063/1.1702424}
  {\bibfield  {journal} {\bibinfo  {journal} {J. Appl. Phys.}\ }\textbf
  {\bibinfo {volume} {33}},\ \bibinfo {pages} {3427} (\bibinfo {year}
  {1962})}\BibitemShut {NoStop}%
\bibitem [{\citenamefont {Jobst}\ \emph {et~al.}(2016)\citenamefont {Jobst},
  \citenamefont {Van Der~Torren}, \citenamefont {Krasovskii}, \citenamefont
  {Balgley}, \citenamefont {Dean}, \citenamefont {Tromp},\ and\ \citenamefont
  {Van Der~Molen}}]{Jobst16}%
  \BibitemOpen
  \bibfield  {author} {\bibinfo {author} {\bibfnamefont {J.}~\bibnamefont
  {Jobst}}, \bibinfo {author} {\bibfnamefont {A.~J.}\ \bibnamefont {Van
  Der~Torren}}, \bibinfo {author} {\bibfnamefont {E.~E.}\ \bibnamefont
  {Krasovskii}}, \bibinfo {author} {\bibfnamefont {J.}~\bibnamefont {Balgley}},
  \bibinfo {author} {\bibfnamefont {C.~R.}\ \bibnamefont {Dean}}, \bibinfo
  {author} {\bibfnamefont {R.~M.}\ \bibnamefont {Tromp}},\ and\ \bibinfo
  {author} {\bibfnamefont {S.~J.}\ \bibnamefont {Van Der~Molen}},\ }\bibfield
  {title} {\bibinfo {title} {Quantifying electronic band interactions in van
  der waals materials using angle-resolved reflected-electron spectroscopy},\
  }\href {https://doi.org/https://doi.org/10.1038/ncomms13621} {\bibfield
  {journal} {\bibinfo  {journal} {Nat. Commun.}\ }\textbf {\bibinfo {volume}
  {7}},\ \bibinfo {pages} {13621} (\bibinfo {year} {2016})}\BibitemShut
  {NoStop}%
\bibitem [{\citenamefont {Krasovskii}(2021)}]{Krasovskii21}%
  \BibitemOpen
  \bibfield  {author} {\bibinfo {author} {\bibfnamefont {E.}~\bibnamefont
  {Krasovskii}},\ }\bibfield  {title} {\bibinfo {title} {Ab initio theory of
  photoemission from graphene},\ }\href {https://doi.org/10.3390/nano11051212}
  {\bibfield  {journal} {\bibinfo  {journal} {Nanomaterials}\ }\textbf
  {\bibinfo {volume} {11}},\ \bibinfo {pages} {1212} (\bibinfo {year}
  {2021})}\BibitemShut {NoStop}%
\bibitem [{\citenamefont {Krasovskii}(2022)}]{Krasovskii22}%
  \BibitemOpen
  \bibfield  {author} {\bibinfo {author} {\bibfnamefont {E.}~\bibnamefont
  {Krasovskii}},\ }\bibfield  {title} {\bibinfo {title} {One-step theory view
  on photoelectron diffraction: Application to graphene},\ }\href
  {https://doi.org/10.3390/nano12224040} {\bibfield  {journal} {\bibinfo
  {journal} {Nanomaterials}\ }\textbf {\bibinfo {volume} {12}},\ \bibinfo
  {pages} {4040} (\bibinfo {year} {2022})}\BibitemShut {NoStop}%
\bibitem [{\citenamefont {Krasovskii}\ and\ \citenamefont
  {Kuzian}(2024)}]{Krasovskii2024}%
  \BibitemOpen
  \bibfield  {author} {\bibinfo {author} {\bibfnamefont {E.~E.}\ \bibnamefont
  {Krasovskii}}\ and\ \bibinfo {author} {\bibfnamefont {R.~O.}\ \bibnamefont
  {Kuzian}},\ }\href@noop {} {\bibinfo {title} {Negative transit time in
  non-tunneling electron transmission through graphene multilayers}} (\bibinfo
  {year} {2024}),\ \bibinfo {note} {to be published}\BibitemShut {NoStop}%
\end{thebibliography}%
\end{document}